\documentstyle[12pt]{article}
\def\Journal#1#2#3#4{{#1} {\bf #2}, #3 (#4)}

\def\NPB{{\em Nucl. Phys.} B}
\def\NPBP{{\em Nucl. Phys.} (Proc. Suppl.) B}
\def\PLB{{\em Phys. Lett.}  B}
\def\PRL{\em Phys. Rev. Lett.}
\def\PRD{{\em Phys. Rev.} D}
\def\ZPC{{\em Z. Phys.} C}
\def\PREP{\em Phys. Rep.}

\newcommand{\titul}[1] {\begin{center}{\large\bf #1 } \end{center}\vskip 1.cm}
\newcommand{\autor}[1] {\begin {center} {\large \lineskip .5em #1 }
                        \end   {center} }
\newcommand{\lugar}[1] {\begin{center} {\it #1} \end{center}}
\newcommand{\abstr}[1] {{\begin{center} \vskip .5cm {\bf Abstract
                        \vspace{0pt}} \end{center}}\begin{quote} #1
                        \end{quote}}

\begin{document}
\begin{titlepage}
.
\begin{flushright} {\bf April 23, 2000} \end{flushright}

\vskip 3.cm
\titul{ On the behavior of $F_2$ \\
and its logarithmic slopes}

\autor{A.B. Kaidalov\footnote{E-mail: kaidalov@vxitep.itep.ru}}
\lugar{ITEP, B. Cheremushkinskaya 25,\\
117259 Moscow, Russia}
\autor{C. Merino\footnote{E-mail: merino@fpaxp1.usc.es}}
\lugar{Departamento de F\'\i sica de Part\'\i culas\\
Universidade de Santiago de Compostela\\
15706 Santiago de Compostela, Spain}
\autor{D. Pertermann\footnote{E-mail: pertermann@physik.uni-siegen.de}}
\lugar{Physics Department, Univ-GH-Siegen\\
D-57068 Siegen, Germany}
\abstr{
It is shown that the CKMT model for the nucleon structure function $F_2$,
taken as the initial condition for the NLO evolution equations in perturbative QCD,
provides a good description of the HERA data when presented in the form 
of the logarithmic slopes of $F_2$ vs $x$ and $Q^2$ (Caldwell-plot), in the whole
available
kinematic ranges. Also the results obtained for the behavior of the gluon
component of a nucleon are presented.
}
\end{titlepage}
\newpage

\pagestyle{plain}

\section{The CKMT model}\indent

The CKMT model \cite{ckmt} for the parametrization of the nucleon structure
function
$F_2$ is a theoretical model based on Regge theory which provides a consistent
formulation of this function in the region of low $Q^2$, and
describes
the
experimental data on
$F_2$ in that region.

The CKMT model \cite{ckmt} proposes
for the nucleon structure functions
\begin{equation}
F_2(x,Q^2) = F_S(x,Q^2) + F_{NS}(x,Q^2),
\label{eq:eq1}
\end{equation}
the following parametrization of its two terms in
the region of small and moderate $Q^2$. For the singlet term, corresponding to
the Pomeron contribution:
\begin{equation}
F_S(x,Q^2) = A\cdot x^{-\Delta(Q^2)}\cdot(1-x)^{n(Q^2)+4}
\cdot\left({Q^2\over Q^2+a}\right)^{1+\Delta(Q^2)},
\label{eq:eq2}
\end{equation}
where the $x$$\rightarrow$0
behavior is determined by an effective intercept
of the Pomeron,~$\Delta$,
which takes into account Pomeron cuts and, therefore (and this is one of the
main points of the model), it depends on $Q^2$. This dependence was
parametrized~\cite{ckmt}
as :
\begin{equation}
\Delta (Q^2) = \Delta_0\cdot\left(1+{\Delta_1\cdot Q^2
\over Q^2+\Delta_2}\right).
\label{eq:eq3}
\end{equation}
Thus, for low values of $Q^2$ (large cuts), $\Delta$ is close
to the effective value found from analysis of hadronic total cross-sections
($\Delta$$\sim$0.08), while for high values of $Q^2$ (small cuts),
$\Delta$ takes the bare Pomeron value,
$\Delta$$\sim$0.2-0.25. The
parametrization for the non-singlet term, which corresponds to the secondary
reggeon (f, $A_2$) contribution, is:
\begin{equation}
F_{NS}(x,Q^2) = B\cdot x^{1-\alpha_R}\cdot(1-x)^{n(Q^2)}
\cdot\left({Q^2\over Q^2+b}\right)^{\alpha_R},
\label{eq:eq4}
\end{equation}
where the $x$$\rightarrow$0 behavior is determined by the secondary
reggeon intercept $\alpha_R$, which is in the range $\alpha_R$=0.4-0.5.
The valence quark contribution can be separated into the contribution of the
u ($B_u$) and d ($B_d$) valence quarks,
the normalization condition for valence quarks fixes
both contributions
at one
given value of $Q^2$ (we use $Q_v^2=2.GeV^2$ in our calculations).
For both the singlet and the non-singlet terms, the behavior when
$x$$\rightarrow$1 is
controlled by $n(Q^2)$, with $n(Q^2)$ being
\begin{equation}
n(Q^2) = {3\over2}\cdot\left(1+{ Q^2
\over Q^2+c}\right),
\label{eq:eq5}
\end{equation}
so that, for $Q^2$=0, the valence quark distributions have the
same power, given by
Regge intercepts, as in the Quark Gluon String Model \cite{kaidalov} or in the Dual
Parton Model \cite{dpm}, $n$(0)=$\alpha_R$(0)$-$$\alpha_N$(0)$\sim$ 3/2, and
the
behavior of
$n(Q^2)$ for large $Q^2$ is taken to coincide with
dimensional counting rules.

The total cross-section for real ($Q^2$=0) photons can be obtained from the
structure function $F_2$ using the following relation:
\begin{equation}
\sigma^{tot}_{\gamma p}(\nu) = \left[{4\pi^2\alpha_{EM}\over Q^2}
\cdot F_2(x,Q^2)\right]_{Q^2=0}.
\label{eq:eq6}
\end{equation}
The proper $F_2(x,Q^2)$$\sim$$Q^2$
behavior when
$Q^2$$\rightarrow$0, is fulfilled in the
model
due to the last factors in equations \ref{eq:eq2} and \ref{eq:eq4}.
Thus, the
$\sigma^{tot}_{\gamma p}(\nu)$ has the following form in
 the CKMT model:
\begin{equation}
\sigma^{tot}_{\gamma p}(\nu) = 4\pi^2\alpha_{EM}
\cdot\left(A\cdot a^{-1-\Delta_0}\cdot(2m\nu)^{\Delta_0}
+(B_u+B_d)\cdot b^{-\alpha_R}\cdot(2m\nu)^{\alpha_R-1}\right).
\label{eq:eq6a}
\end{equation}

The parameters were determined \cite{ckmt} from a joint fit of the
$\sigma^{tot}_{\gamma p}$ data and the NMC data \cite{nmc} on
the proton structure function in the region
$1 GeV^2 \leq Q^2 \leq 5 GeV^2$, and a very good
description of the experimantal data available was obtained.

The next step in this approach is to introduce the QCD evolution in
the partonic distributions of the CKMT model and thus to
determine the structure
functions at higher values of $Q^2$. For this, the
evolution equation in two loops in the $\overline{\mbox{MS}}$
scheme with
$\Lambda=200.MeV$ was used \cite{ckmt}.

The results obtained by taking into account the QCD evolution in
this way are \cite{ckmt} in a very good agreement with
the experimental data on
$F_2(x,Q^2)$ at high values of $Q^2$.

When the publication of the
data \cite{h1,zeus} on $F_2$ from HERA at low and
moderate $Q^2$ provided the opportunity to include in the fit of
the parameters
of the model experimental points from the kinematical
region where the CKMT parametrization should give a good description
without need of
any perturbative QCD evolution,
one proceeded~\cite{newpaper} to add
these
new data on $F_2$ from H1 and ZEUS at low and moderate $Q^2$, to
those from NMC \cite{nmc} and E665 \cite{e665} collaborations, 
and to data \cite{cross}
on
cross-sections for real photoproduction, into a global fit which
allowed
the test
of the model in wider regions of $x$ and $Q^2$. One took as
initial condition
for the values of the different parameters
those obtained in the previous fit \cite{ckmt}, and
although the quality of the fit is not very sensitive to small 
changes in the values of the parameters, the best fit has been found
for the values of the parameters given in Table\ref{tab:tab1}.
\begin{table}[h]
\caption{Values of the parameters in the CKMT model obtained in
the fit of $F_2$ when also the low $Q^2$ HERA data
are included.
All dimensional parameters are given
in $GeV^2$. The
valence counting rules provide the following values of $B_u$ and $B_d$, for the
proton case, when fixing their normalization at $Q_v^2$=2.$GeV^2$: 
$B_u$=1.1555, $B_d$=0.1722.
\label{tab:tab1}}
\vspace{0.2cm}
\centering
\vskip 1.cm
\hskip 1.cm
\footnotesize
\vbox {\offinterlineskip
\hrule
\halign{&\vrule#&
\strut\quad\hfil#\quad\cr
height2pt&\omit&&\omit&\cr
&CKMT model\hfil&&values of the parameters&\cr
height2pt&\omit&&\omit&\cr
\noalign{\hrule}
height2pt&\omit&&\omit&\cr
&A\hfil&&0.1301\hfil&\cr
height2pt&\omit&&\omit&\cr
\noalign{\hrule}
height2pt&\omit&&\omit&\cr
&a\hfil&&0.2628\hfil&\cr
height2pt&\omit&&\omit&\cr
\noalign{\hrule}
height2pt&\omit&&\omit&\cr
&$\Delta_0$\hfil&&0.09663\hfil&\cr
height2pt&\omit&&\omit&\cr
\noalign{\hrule}
height2pt&\omit&&\omit&\cr
&$\Delta_1$\hfil&&1.9533\hfil&\cr
height2pt&\omit&&\omit&\cr
\noalign{\hrule}
height2pt&\omit&&\omit&\cr
&$\Delta_2$\hfil&&1.1606\hfil&\cr
height2pt&\omit&&\omit&\cr
\noalign{\hrule}
height2pt&\omit&&\omit&\cr
&c\hfil&&3.5489 (fixed)\hfil&\cr
height2pt&\omit&&\omit&\cr
\noalign{\hrule}
height2pt&\omit&&\omit&\cr
&b\hfil&&0.3840\hfil&\cr
height2pt&\omit&&\omit&\cr
\noalign{\hrule}
height2pt&\omit&&\omit&\cr
&$\alpha_R$\hfil&&0.4150 (fixed)\hfil&\cr
height2pt&\omit&&\omit&\cr}
\hrule}
\end{table}
\vspace{0.5cm}

The
quality of the
description 
provided by the
CKMT model
of all the experimental data 
on $\sigma^{tot}_{\gamma p}$ and $F_2$,
and, in
particular, of the
the new experimental data from HERA is very high, with a value
of $\chi^2/d.o.f.$ for the global fit, $\chi^2/d.o.f.$=106.95/167,
where
the statistical and
systematic
errors have been treated in quadrature, and where the relative normalization
among all the experimental data sets has been taken equal to 1.

Thus, by taking into account the general features of the CKMT model
described above, we use the CKMT model
to describe
the experimental
data in the region of low $Q^2$
($0< Q^2< Q^2_0$),
and
then we take this parametrization as the initial condition at
$Q^2_0$ to be
used in the NLO QCD evolution equation to obtain 
$F_2$ at values of $Q^2$ higher than $Q^2_0$. In order to determine the distributions 
of gluons in a nucleon the CKMT model assumes \cite{ckmt} that the only difference
between distributions of sea-quarks and gluons is in the $x\rightarrow 1$ 
behavior. Following \cite{fmartin} we write it in the form
\begin{equation}
xg(x,Q^2)=Gx\bar{q}(x,Q^2)/(1-x),
\label{neweq1}
\end{equation}
where $x\bar{q}(x,Q^2)$ is proportional to the expression in equation~\ref{eq:eq2}.
The constant $G$ is determined from the energy-momentum conservation sum rule.
 
We have performed 
our calculations for two different values of $Q_0^2=2.GeV^2$ and $Q_0^2=4.GeV^2$.
We also show our results in the shape of both the $dF_2/dlnQ^2$ and the
$dlnF2/dln(1/x)$ slopes in order to compare with the experimental
data when given in the so-called Caldwell-plot.
This approach provides a smooth transition from the region of small
$Q^2$, which is governed by the physics of Regge theory, to a 
region of large $Q^2$, where the effects of QCD-evolution are
important.

The way we proceed to calculate $F_2$, and its 
logarithmic derivatives
$dF_2/dlnQ^2$, and $dlnF2/dln(1/x)$, is the following (see
Appendix A for all the technical details
on how the NLO QCD evolution has been performed):
\begin{itemize}

\item  In the region $0< Q^2\le Q^2_0$ we use the pure
CKMT model for $F_2$.

\item  For $Q^2_0< Q^2\le charm$ $threshold$ \cite{grv},
we make the QCD
evolution of $F_2$ at NLO in the $\overline{\mbox{MS}}$
scheme for a
number of flavours $n_f=3$, and we take
as the starting parametrization that given by the
CKMT model. For $Q^2_0$ we have used in this calculation 
two different values:
$Q^2_0=2.GeV^2$, and $Q^2_0=4.GeV^2$.

\item  When $charm$
$threshold < Q^2\le \bar{Q}^2=50. GeV^2$, also
the QCD evolution of $F_2$ is implemented at NLO in the
$\overline{\mbox{MS}}$
scheme for a
number of flavours $n_f=3$, using the parton distribution
functions for the $u,d,s$ quarks, and by including the charm
contribution via photon-gluon fusion.

\item  For values of $Q^2> \bar{Q}^2$, QCD evolution is computed
at NLO in the $\overline{\mbox{MS}}$
scheme, but now with a
number of flavours $n_f=4$, and by using the parton distribution
functions for the $u,d,s$, and $c$ quarks.

\end{itemize}

One has to note that in the treatment of the charm contribution we
have followed reference \cite{haakman}.

\section{ Results} \indent

The results we have obtained are presented in figures
1 to 9.

Figure 1 shows $F_2(x,Q^2)$ as a function of $x$ for 
several values of $Q^2$, from 
$Q^2=0.6 GeV^2$ to $Q^2=17. GeV^2$. The dotted lines correspond 
to the pure CKMT model without any perturbative evolution, while the full
lines run for the evoluted CKMT parametrization. When for a given
value of $Q^2$ two full lines are depicted, the bold (solid) one has been
obtained by taking the starting point for the QCD evolution as
$Q_0^2=2. GeV^2$ ($Q_0^2=4. GeV^2$). Experimental
points in this figure are from E665 \cite{e665}, H1 \cite{H1caldwell}, 
and ZEUS \cite{caldw} collaborations.

In Figures 2.a and 2.b, we present the comparison 
of the pure CKMT 
parametrization of $F_2$ with the low $Q^2$ data of E665, 
ZEUS-BPC95, and ZEUS-BPT97, as compiled in \cite{amelung} and \cite{zhokin}.
One sees that the agreement between the CKMT model and the experimental data
in this region of low $Q^2$ is good. 

In Figure 3 (Caldwell-plot), the slope
$dF_2/dlnQ^2$ is shown as a function of $x$, and compared with
the $a+blnQ^2$ fit to the ZEUS $F_2$ data in bins of $x$.
This plot was considered as an evidence for a transition from
hard to soft regime of QCD in the region of $Q^2\sim 5.GeV^2$
(see for example \cite{amueller}). This question has been studied 
theoretically in references \cite{levin,jenkovszky}. Figure 3
shows that the CKMT model is in a good agreement with experimental
points in the whole region of $x$ and $Q^2$. One problem with the
presentation of the data in Figure 3 is a strong correlation
between $x$ and $Q^2$ values for the data points. It follows from the 
formulas of CKMT model for $dF_2(x,Q^2)/dlnQ^2$ given in Appendix B
that for a fixed value of $Q^2$ this quantity monotonically
increases as $x\rightarrow 0$. The existence of a maximun
of $dF_2(x,Q^2)/dlnQ^2$ in Figure 3 is related
to the correlation between $Q^2$ and $x$ in the region of small
$x$ (or $Q^2$). The same conclusion was achieved in reference
\cite{levin}, and recently confirmed by experimental
data \cite{zhokin}.  

Figures 4 and 5 show the slope
$dlnF2/dln(1/x)$ as a function of $Q^2$ compared to the fits
$F_2=Ax^{-\Delta_{eff}}$ of the the ZEUS \cite{caldw} and 
H1 \cite{H1caldwell} data,
respectively. In Figure 4, as the $x$ range of the BPC95
data is restricted, also the E665 \cite{e665} data
were included in \cite{caldw}, and are now also taken into account.
This slope is sometimes interpreted as
the $\Delta_{eff}$ of the
Pomeron exchange, $\Delta_{eff}=dlnF2/dln(1/x)$.
Let us note that in our approach $\Delta_{eff}$ for
$Q^2 > Q^2_0$ can not be interpreted as an effective Pomeron 
intercept, because the QCD evolution leads to a substantial increase
of $\Delta_{eff}$ as $Q^2$ increases. On the other hand this effect
should decrease as $x\rightarrow 0$.

In the experimental fits, each $Q^2$ bin corresponds to
a average value of $x$, $<x>$, calculated from the mean value of
$ln(1/x)$ weighted by the statistical errors of the corresponding
$F_2$ values in that bin. Even though we can proceed as in the
experimental fits, and we get a very good agreement with the data,
since the estimation of $<x>$ is in some
sense artificial and arbitrary, and it introduces unphysical wiggles
when drawing one full line connecting the different bins,
we made for all the $Q^2$ bins in this figures the choice of the
smallest $x$ in the data,
instead of considering a different $<x>$
for each $Q^2$. This choice is based
on the fact that the ansatz $\Delta_{eff}=dlnF2/dln(1/x)$ is
actually valid for small $x$,
and it results in a smooth curve except for the
jump in the region around $Q^2\sim 50.GeV^2$, where the evolution
procedure changes (again, see Appendix A for more details).

Since the structure function $F_2$ in the region of
low $x$ is determined at large extent by the gluon component, we
present our prediction for the behavior of this gluon component.
Thus, Figure 6 shows the gluon density distribution as a function of
$Q^2$ calculated by performing the NLO QCD evolution of the CKMT
model, and its comparison with the H1 Collaboration data 
in reference \cite{adloff}, and Figure 7 represents the gluon
densities at $\mu^2=25.GeV^2$ as a function of $x$ calculated by 
evoluting the CKMT model at NLO in the QCD evolution, and
compared to those determined from H1 DIS and photoproduction 
data. Experimental data on $D^*$ meson cross-section measurements
are from references \cite{zhokin,adloff}.
Figure 8 shows the behavior of $xg(x,\mu^2)$ at 
$\mu^2=200~GeV^2$ as a function of $Q^2$ to be
compared with the H1-dijets results \cite{zhokin,wobisch}.
Finally, Figure 9 shows the prediction
of the CKMT model for $xg(x,Q^2)$ as a function of $x$ at the values 
of $Q^2$ measured both by H1 and ZEUS collaborations.

A satisfactory agreement with the experiment is obtained in the
whole ranges of
$x$ and $Q^2$ where experimental data are available, 
showing that the experimental behavior of $F_2$, its logarithmic
slopes, and its gluon component can be
described by using as initial condition for the QCD evolution equation
a model of $F_2$ where the shadowing effects which are important at low
values of $Q^2$ are included, like the CKMT model.

\section{Conclusions} 
The CKMT model for the parametrization of
the nucleon
structure functions
provides a very good description of all the
available experimental data on $F_2(x,Q^2)$ at low and moderate $Q^2$,
including the 
recent small-$x$ HERA points.

An important ingredient of the model is the dependence of an
effective intercept of the Pomeron on $Q^2$. It has been shown
recently \cite{elenagf} that such a behavior is naturally reproduced
in a broad class of models based on reggeon calculus, which describes
simultaneously the structure function $F_2$ and the diffractive
production by virtual photons.

Use of the CKMT model as the initial condition for the QCD-evolution
equations in the region of $Q^2=2.\div5.GeV^2$ leads to a good 
description of all available data in a broad region of $Q^2$, 
including the logarithmic slopes of the structure function
$F_2(x,Q^2)$, $dF_2(x,Q^2)/dlnQ^2$ and 
$dlnF_2(x,Q^2)/dln(1/x)$. Thus an unified description of the data
on $F_2$ for all values of $Q^2$ is achieved.


\vspace{1cm}
\hspace{1cm} \Large{} {\bf Acknowledgements}    \vspace{0.5cm}

\normalsize{}

A.B.K. acknowledges support of a NATO grant 
OUTR.LG971390 and the RFBR grant 98-02-17463,
and C.M. was partially supported by
CICYT (AEN99-0589-C02-02).

\newpage
\vspace{0.5cm}
\hspace{1cm} \Large{} {\bf Appendix A  -- NLO QCD evolution of $F_2(x,Q^2)$}    
\vspace{0.5cm}
\normalsize{} 

For the reader convenience we present here some technical remarks concerning 
the  NLO QCD calculation of $F_2(x,Q^2)$.

For sufficiently large $Q^2 >1~GeV^2$,  the structure function $F_2(x,Q^2)$ can 
be expressed by perturbative parton distributions. In leading order (LO) perturbation 
theory, the expression is given as
\begin{equation}
  \label{apx01}
  \frac{1}{x}F_2(x,Q^2) = x \sum_q  e_q^2  \{ q(x,Q^2) + \bar{q}(x,Q^2) \},
\end{equation}
where $q$ and $\bar{q}$ denote the quark and anti-quark distribution functions, 
$e_q^2$ the square of the quark electric charge, and the sum runs over all quark 
flavors included \cite{grv}.  On the other hand, with $F_2(x,Q^2)$ given 
in eqs. (\ref{eq:eq1}-\ref{eq:eq5}), and making reasonable
assumptions concerning the flavor structure of the QCD-sea, one 
can extract from $F_2(x,Q^2)$ the different parton distribution functions, 
including that of the gluon
component \cite{ckmt}. Generally, the calculation of $F_2(x,Q^2)$  at 
$Q^2 \gg 1~GeV^2$ requires a $Q^2$-evolution \`a la DGLAP \cite{DGLAP00}. 
The procedure consists in the solution of the LO-DGLAP equations for the parton
distribution functions using reasonable initial distributions at a starting value 
$Q^2=Q_0^2$  ($1~GeV^2 < Q_0^2 < 5~GeV^2$).  Using eq.(\ref{apx01}),
the resulting quark distributions at $Q^2$ can be recombined to $F_2$ at this 
virtuality. 
   
By the evolution of the CKMT-model we mean the application of this procedure to the model
discussed in this paper. As mentioned above, the CKMT-model of $F_2(x,Q^2)$ 
is valid within $0 \le Q^2 < 5~GeV^2$. Due to the good aggreement with 
experimental data the parton distributions extracted from $F_2^{CKMT}$ at a  $Q_0^2$ 
in the range given above seem to be reasonable initial distributions for an 
evolution to higher $Q^2$.

In next to leading order (NLO), the relation between $F_2(x,Q^2)$ and the parton
distribution functions is more complicated and depends on the renormalization scheme. 
The calculations presented here are performed in the $\rm \overline{MS}$-scheme 
\cite{Bardeen:1978yd}. 
In this context, the structure function is given by \cite{grv} as
\begin{equation}
 \label{apx02}
  \frac{1}{x}F_2(x,Q^2) = \sum_q e_q^2 
  \left\{ q(x,Q^2)+\bar{q}(x,Q^2) + \frac{\alpha_s(Q^2)}{2\pi}
  \left[ C_{q,2}*(q+\bar{q}) + 2\cdot C_{g,2}*g \right] 
  \right\},
\end{equation}
where $q,\bar{q}$ and $g$ are the NLO quark, anti-quark and gluon distribution functions,
respectively.  $\alpha_s$ denotes the strong coupling constant in NLO. The convolutions 
$C*q$ and $C*g$ are defined as
\begin{equation}
  \protect\label{apx03}
  C*q = \int_x^1 \frac{dy}{y} C \left( \frac{x}{y} \right) q(y,Q^2).
\end{equation}
The Wilson coefficients $C_{q,g,2}(z)$ are given by
\begin{eqnarray}
  \protect\label{apx04}
 C_{q,2}(z) = \frac{4}{3}\left[ \frac{1+z^2}{1-z}\left( \ln\frac{1-z}{z}
              - \frac{3}{4} \right) + \frac{1}{4}(9+5z) \right]_+, && 
 \nonumber \\
 C_{g,2}(z) = \frac{1}{2}\left[ (z^2 + (1-z)^2)\ln\frac{1-z}{z}
              - 1 + 8z(1-z) \right].
\end{eqnarray}
Here, the  integral over a $[\cdot]_+$-distribution is defined as described in \cite{grv2}:
\begin{eqnarray}
  \protect\label{apx05}
  C_+*q &=& \int_x^1 \frac{dy}{y} C \left( \frac{x}{y} \right)_+ q(y,Q^2)
  \nonumber \\
        &=& \int_x^1 \frac{dy}{y} C \left( \frac{x}{y} \right)
            \left[ q(y,Q^2) - \frac{x}{y} q(x,Q^2)\right]
            - q(x,Q^2) \int_0^x dy C(y).
\end{eqnarray}

There are alternative renormalization schemes as, for instance, the $DIS$-scheme \cite{grv}.
Here, the form of eq. (\ref{apx01}) is kept for NLO also, i.e. 
\begin{equation}
  \label{apx06}
  \frac{1}{x}F_2(x,Q^2) = x \sum_q  e_q^2  \{ q_{DIS}(x,Q^2) + \bar{q}_{DIS}(x,Q^2) \}.
\end{equation}
The relation between the $\overline{MS}$- and  the $DIS$-distributions is given by
\begin{eqnarray}
  \protect\label{apx07}
  \stackrel{(-)}{q}_{DIS}(x,Q^2) &=& \stackrel{(-)}{q}(x,Q^2) +
  \frac{\alpha_s(Q^2)}{2\pi}\left[ C_{q,2}*\stackrel{(-)}{q} +
  C_{g,2}*g \right] +
  O(\alpha_s^2),
\\
  g_{DIS}(x,Q^2) &=& g(x,Q^2) - \frac{\alpha_s(Q^2)}{2\pi}
  \left[\sum_q C_{q,2}*(q+\bar{q}) + 2f\cdot C_{g,2}*g \right] +
  O(\alpha_s^2). \nonumber
\end{eqnarray}
The parameter $f$ denotes the number of active flavors in the sea.

Our procedure to extract the parton-distributions from $F_2^{CKMT}$ is based on the 
LO-formula eq. (\ref{apx01}).  
Therefore,  in NLO, we extract the $DIS$-distributions. 
Now, the task is to calculate the $\overline{MS}$-distributions at $Q^2=Q_0^2$ .
This can be done using a first order approximation in $\alpha_s(Q^2)/2\pi$:
\begin{eqnarray}
  \protect\label{apx08}
  \stackrel{(-)}{q}(x,Q_0^2) &\approx& \stackrel{(-)}{q}_{DIS}(x,Q_0^2) -
  \frac{\alpha_s(Q_0^2)}{2\pi}\left[ C_{q,2}*\stackrel{(-)}{q}_{DIS} +
  C_{g,2}*g_{DIS} \right]
\\
  g(x,Q_0^2) &\approx& g_{DIS}(x,Q_0^2) + \frac{\alpha_s(Q_0^2)}{2\pi}
  \left[\sum_q C_{q,2}*(q_{DIS}+\bar{q}_{DIS}) + 
  2f\cdot C_{g,2}*g_{DIS} \right].
  \nonumber
\end{eqnarray}

In summary, the $Q^2$-evolution of $F_2^{CKMT}$  works as follows:
\begin{enumerate}
\item One chooses an appropriate value $Q^2=Q_0^2 > 1~GeV^2$ as a starting point for the 
           evolution. These are $Q_0^2= 2~GeV^2$ or $Q_0^2= 4~GeV^2$ in our calculations.
\item At $Q^2=Q_0^2$, one extracts the NLO parton distribution functions from  
           $F_2^{CKMT}$ . The relation between these parton distributions and the structure 
           function is given by eq.(\ref{apx06}) which is formally the same as eq.(\ref{apx01}) 
           in LO. So the resulting parton distributions are the $DIS$-functions, i.e. 
           $q_{DIS}(x,Q_0^2),~\bar{q}_{DIS}(x,Q_0^2)$ and $g_{DIS}(x,Q_0^2)$.
\item Using eq.(\ref{apx08}) one calculates the $\overline{MS}$-distributions 
           $q(x,Q_0^2),~\bar{q}(x,Q_0^2)$ and $g(x,Q_0^2)$.
\item These $\overline{MS}$-functions serve as initial distributions in a numerical  procedure 
           to solve the NLO-DGLAP-equations in the $\overline{MS}$-scheme for a certain value 
           $Q^2 > Q_0^2$. The result are the evoluted $\overline{MS}$-parton distributions  
           $q(x,Q^2),~\bar{q}(x,Q^2)$ and $g(x,Q^2)$. 
\item Finally, using eq.(\ref{apx02}), the structure function $F_2^{CKMT}(x,Q^2)$ can be 
           recalculated.
\end{enumerate}

The charm production is of particular interest. Following refs. \cite{grv,haakman}, 
the assumption of a ``massless'' charm quark produced above the threshold  $Q^2_c=4m_c^2$ 
($m_c^2$ -- charm quark mass) via the usual DGLAP-evolution is not realistic. 
This procedure is useful in the range of high $Q^2 \gg Q_c^2$ only. In the intermediate
region $Q_c^2 < Q^2 < \bar{Q^2}=50~GeV^2$, the charm is treated via a photon-gluon 
fusion process. The corresponding contribution to the structure function is defined as 
\begin{equation}
  \label{apx20}
  \frac{1}{x} F_2^c(x,Q^2,m_c^2) ~=~ 
  2e_c^2~\frac{\alpha_s(\mu^2)}{2\pi} \int_{ax}^1\frac{dy}{y}~\cdot
  C_{g,2}^c \left( \frac{x}{y},\frac{m_c^2}{Q^2} \right)  \cdot g(y,\mu^2),
\end{equation}
where $\mu^2=4m_c^2$, $a=1~+~4m_c^2/Q^2$, and,
the coefficient $C_{g,2}^c (Z,R)$ is given by
\begin{eqnarray}
  \label{apx21}
  C_{g,2}^c(Z,R) & = &
  \frac{1}{2} 
  \Big\{  [Z^2 + (1-Z)^2 + 4ZR(1-3Z) -8Z^2R^2] 
  \ln{\frac{1+V}{1-V}}  \nonumber \\
  &+&V [-1 + 8Z(1-Z) - 4ZR(1-Z)] \Big\},
  \end{eqnarray}
with $V^2=1-4RZ/(1-Z)$. So, $F_2$ in total is given by eq.(\ref{apx02}) where the sum
runs over $q=u,d,s$ plus eq.(\ref{apx20}). The contributions of the botom and top quarks are
neglected here. Precisely, the charm threshold is defined as discussed in 
refs. \cite{grv,haakman},
\begin{equation}
  \label{apx23}
  W^2 \equiv Q^2 (1/x~-~1) \ge Q_c^2 = 4m_c^2.  
\end{equation}

The $Q^2$-dependence of $F_2$ can be summarized as follows:
\begin{description}
\item[i) $Q^2 < Q_0^2$:] In the low $Q^2$ region, $F_2$ is calculated as given in
                                                            the pure CKMT-model, eqs.(\ref{eq:eq1}--\ref{eq:eq5}).
\item[ii) $Q_0^2 < Q^2 < Q_c^2$:] Below the charm  threshold  $F_2$ is calculated
         using eq.(\ref{apx02}) from NLO QCD evoluted parton distributions in the 
        $\overline{MS}$-scheme. The number of flavors is $f=3$ ($u,d,s$). 
\item[iii) $Q_c^2 < Q^2 < \bar{Q}=50~GeV^2$:] Above the charm  threshold  $F_2$ 
          is determined from eqs.(\ref{apx02}) and (\ref{apx20}). Note that the number of
          flavors active in the evolution is again $f=3$ ($u,d,s$). However,  $f=4$ after the charm is
          produced. This is important for the calculation of $\alpha_s$. 
\item[iv) $Q^2 > \bar{Q}^2=50~GeV^2$:] In the high $Q^2$-region, $F_2$ is given
          by eq.(\ref{apx02}). The charm is produced as ``massless''  quark in the evolution
          process. Generally, the number of flavors is $f=4$. 
\end{description}
The threshold $\bar{Q}^2$ where the charm production in the evolution process is more 
important than the photon-gluon fusion is discussed in detail in \cite{haakman}. The value
of $50~GeV^2$ is chosen to guarantee a transition as smooth as possible. This method works
better for $x \rightarrow 0$ than for $x \rightarrow 1$. This explains the small wiggles in some 
of the figures at $Q^2=50~GeV^2$.

\newpage
\vspace{0.5cm}
\hspace{1cm} \Large{} {\bf Appendix B  -- The slopes of $F_2(x,Q^2)$}    
\vspace{0.5cm}
\normalsize{} 

For low $Q^2$, the ``pure''  CKMT-model is used, i.e. the one defined in 
eqs.(\ref{eq:eq1}-\ref{eq:eq5}). Here, the calculation of the slopes as
$dF_2(x,Q^2)/d\ln{Q^2}$ and $d\ln{F_2}(x,Q^2)/d\ln{(1/x)}=\Delta_{eff}$ is 
straightforward. Considering $x$ and $Q^2$ as independent variables one gets
\begin{equation}
\begin{array}{rcl}
\frac{\textstyle dF_2(x,Q^2)}{\textstyle dlnQ^2} =& F_S(x,Q^2) & 
[\frac{\Delta_2}{Q^2+\Delta_2}
\left(\Delta(Q^2)-\Delta_0\right)
ln\frac{Q^2}{x(Q^2+a)}\\
&&
+\frac{c}{Q^2+c}
\left(n(Q^2)-\frac{3}{2}\right)ln(1-x)
+\frac{a~\left(1+\Delta(Q^2)\right)}{Q^2+a}]\\
+ &F_{NS}(x,Q^2)& 
[\frac{c}{Q^2+c}\left(n(Q^2)-\frac{3}{2}\right)ln(1-x)
+\frac{b~\alpha_R(0)}{Q^2+b}],\\ & &
\end{array}
\label{apx09}
\end{equation}
which in the limit $Q^2\rightarrow 0$ takes the form
\begin{equation}
\begin{array}{rcl}
\frac{\textstyle dF_2(x,Q^2)}{\textstyle dlnQ^2} & \sim &
\left(1+\Delta_0\right) F_S(x,Q^2) + \alpha_R(0) F_{NS}(x,Q^2).
\end{array}
\label{apx10}
\end{equation}

Also, if one considers the case when W is fixed one can take $x\sim C\cdot Q^2$,
and then, up to constant factors, one gets:
\begin{equation}
\begin{array}{rcl}
\frac{\textstyle dF_2(x,Q^2)}{\textstyle dlnQ^2} =& F_S(x,Q^2) &
[-\frac{\Delta_2}{Q^2+\Delta_2}\left(\Delta(Q^2)-\Delta_0\right)
ln(Q^2+a)\\
&&
-\Delta(Q^2)+\frac{c}{Q^2+c}\left(n(Q^2)-\frac{3}{2}\right)ln(1-Q^2)\\
&&
-\frac{Q^2n(Q^2)}{1-Q^2}+\frac{a~(1+\Delta(Q^2))}{Q^2+a}]\\
+ & F_{NS}(x,Q^2) &
[\frac{c}{Q^2+c}(n(Q^2)-\frac{3}{2})ln(1-Q^2)\\
&&
+\frac{b~\alpha_R(0)}{Q^2+b}+(1-\alpha_R(0))-\frac{Q^2n(Q^2)}{1-Q^2}].\\
&&
\end{array}
\label{apx11}
\end{equation}
Now, if one takes W fixed with $Q^2\sim x\rightarrow 0$,
one can easily see
that this equation simply reduces to:
\begin{equation}
\frac{dF_2(x,Q^2)}{dlnQ^2} \sim F_2(x,Q^2).
\label{apx12}
\end{equation}

The calculations presented in the paper are based on the assumption of independent 
$x$ and $Q^2$, i.e. eqs.(\ref{apx09}, \ref{apx10}). In this context, the effective x-slope
$\Delta_{eff}=d\ln{F_2(x,Q^2)}/d\ln{(1/x)}$ is given by
\begin{equation}
\begin{array}{rl}
F_2(x,Q^2)  \cdot
\frac{\textstyle d\ln{F_2(x,Q^2)}}{\textstyle d\ln{(1/x)}} =& 
[\Delta(Q^2) + \frac{x}{1-x} (n(Q^2)+4)] \cdot F_S\\
+&[\alpha_R(0) - 1 + \frac{x}{1-x} n(Q^2) + \frac{x~B_d}{B_u + B_d(1-x)}]
\cdot F_{NS}.
\end{array}
\label{apx13}
\end{equation}

For $Q^2 > Q_0^2$,  the slopes have to be calculated from the evoluted structure function. 
Here, there are two fundamental procedures, the pure numerical and the mainly analytical 
calculations.  The pure numerical procedure is very simple:
\begin{equation}
\begin{array}{rl}
\frac{\textstyle dF_2(x,Q^2)}{\textstyle d\ln{Q^2}} \approx & 
Q^2 \cdot
\frac{1}{2\delta Q^2} \cdot [F_2(x,Q^2 + \delta Q^2) - F_2(x,Q^2 - \delta Q^2)], 
\end{array}
\label{apx14}
\end{equation}
\begin{equation}
\begin{array}{rl}
\frac{\textstyle d\ln{F_2(x,Q^2)}}{\textstyle d\ln{(1/x)}} \approx & 
(-1)\cdot \frac{\textstyle x}{\textstyle F_2(x,Q^2)} \cdot
\frac{1}{2\delta x} \cdot [F_2(x + \delta x,Q^2) - F_2(x - \delta x,Q^2)]. 
\end{array}
\label{apx15}
\end{equation}
$F_2(x,Q^2)$ is the evoluted structure function whereas $\delta Q^2$ and $\delta x$ denote
the corresponding increments which are fixed to be $10^{-3}\cdot Q^2$ or
 $10^{-3}\cdot x$ in the  calculations presented. For low $Q^2$, we have checked this 
procedure comparing the values of eqs. (\ref{apx14},\ref{apx15}) with those calculated using 
eqs. (\ref{apx09},\ref{apx13}).  The agreement is very good which, in some cases, is 
demonstrated by  identical numbers.
This numerical procedure is the method used to determine  the effective x-slope
$\Delta_{eff}=d\ln{F_2(x,Q^2)}/d\ln{(1/x)}$ of the evoluted structure function.
In the case of  $dF_2(x,Q^2)/d\ln{Q^2}$ there is, in addition, the way of mainly analytical 
calculations. If the parton distribution functions are known their derivatives concerning 
$Q^2$ can be calculated from the DGLAP-equations. Instead of $Q^2$ the parameter
\begin{equation}
\begin{array}{rcl}
S = \ln \left\{ \frac{T}{T_o} \right\}, &
T = \ln(Q^2/\Lambda_{QCD}^2), &
T_o = \ln(Q_0^2/\Lambda_{QCD}^2)
\end{array}
\label{apx16}
\end{equation}
is often used in perturbation theory. In terms of $S$
\begin{equation}
\frac{dF_2}{d\ln{Q^2}} = 
\frac{1}{\ln(Q^2/\Lambda_{QCD}^2)} \frac{dF_2}{dS},
\label{apx17}
\end{equation} 
and in the $\overline{MS}$-scheme
\begin{eqnarray}
  \protect\label{apx18}
  \frac{1}{x} \frac{dF_2(x,S)}{dS} &=& \sum_q e_q^2 
  \Bigg\{ \frac{dq(x,S)}{dS}+\frac{d\bar{q}(x,S)}{dS}
  \nonumber \\
  && + \frac{\alpha_s(Q^2)}{2\pi}
  \left[ C_{q,2}*(\frac{dq}{dS}+\frac{d\bar{q}}{dS}) + 
  2\cdot C_{g,2}*\frac{dg}{dS} \right] 
  \nonumber \\
  && + \frac{1}{2\pi}\frac{d\alpha_s(Q^2)}{dS}
  \left[ C_{q,2}*(q+\bar{q}) + 2\cdot C_{g,2}*g \right] 
  \Bigg\}.
\end{eqnarray}

The numerical integration procedure for solving the DGLAP-equations used in the work
presented here gives the evoluted parton distributions and their derivatives on $S$ as
the output. In NLO, $d\alpha_s(Q^2)/dS$ is simple to calculate,
\begin{eqnarray}
  \protect\label{apx19}
  \frac{\alpha_s(T)}{2\pi} & = & \frac{2}{\beta_0T}\left(1 - 
                \frac{\beta_1}{\beta_0^2}\frac{\ln(T)}{T}
                \right), \nonumber \\
 \frac{1}{2\pi}\frac{d\alpha_s(T)}{dT} &=& -\frac{1}{T}\cdot
                 \frac{\alpha_s(T)}{2\pi} + \frac{2\beta_1}{\beta_0^2 T^3}(\ln(T)-1),\\
 \frac{1}{2\pi}\frac{d\alpha_s}{dS} & = &
                 T\cdot  \frac{1}{2\pi}\frac{d\alpha_s}{dT}.  \nonumber 
\end{eqnarray}
Thus, with the derivatives $dq/dS$, $d\bar{q}/dS$ and $dg/dS$ one gets the 
$Q^2$-derivative of $F_2$. This method is called as ``mainly analytical''
(it includes a numerical integration procedure).

Eq.(\ref{apx18}) is valid below the charm threshold \cite{grv,haakman}, 
i.e. $Q_0^2 < Q^2 < Q_c^2$, and
in the high $Q^2$-region where the charm can be considered as a ``massless'' dynamical 
quark \cite{haakman}. As described above, the charm
is treated via a photon-gluon fusion process in the range 
$Q_c^2 < Q^2 < \bar{Q^2}=50~GeV^2$ \cite{grv,haakman}. 
From eq. (\ref{apx20}) the charm slope contribution can be detrmined as
\begin{equation}
  \label{apx22}
  \frac{1}{x} \frac{dF_2^c(x,Q^2,m_c^2)}{d\ln{Q^2}} ~=~ 
  2e_c^2~\frac{\alpha_s(\mu^2)}{2\pi} \int_{ax}^1\frac{dy}{y}~\cdot
  \frac{dC_{g,2}^c}{dlnQ^2} \left( \frac{x}{y},\frac{m_c^2}{Q^2} \right) 
  \cdot g(y,\mu^2).
\end{equation}
The total slope is the sum of eqs. (\ref{apx18}) and (\ref{apx22}).

We have calculated the $Q^2$-slope of the evoluted $F_2$ in the perturbative 
region \newline ($Q^2 \ge Q_0^2$) using both, the numerical and the analytical methods.
The values are in agreement although the difference increases somewhat in the
region near to $Q_0^2$. The values presented in the figures are from the numerical
calculation.

\newpage

\vspace{0.5cm}

\hspace{1cm} {\Large{\bf Figure captions}}    
\vspace{0.5cm}

{\bf Figure 1.} 
$F_2$ as a function of $x$ computed  
in the CKMT model 
for twelve different values of $Q^2$, and compared with the following
experimental data (see \cite{caldw} for the experimental references):
ZEUS SVX95 (black circles), H1 SVX95 (white
triangles), ZEUS BPC95 (white squares), E665 (white diamonds), and ZEUS
94 (white circles). The dotted line is the theoretical result obtained
with the pure CKMT model, and the bold (solid) line is the result 
obtained with
the NLO QCD-evoluted CKMT model when one takes $Q^2_0=2. GeV^2$
($Q^2_0=4. GeV^2$).

\vspace{0.5cm}

{\bf Figure 2.} 
$F_2$ as a function of $x$ computed in the CKMT model
for six (a) and five (b) different low values of $Q^2$, and compared 
with the following
experimental data (see \cite{caldw,amelung,zhokin} 
for the experimental references):
ZEUS BPT97 (black circles), 
ZEUS BPC95 (white circles), and E665 (white squares).
The theoretical result has been obtained
with the pure CKMT model.

\vspace{0.5cm}

{\bf Figure 3.} 
$dF_2/dlnQ^2$ as a function of $x$ computed
by performing the NLO QCD perturbative evolution of the 
CKMT model (see appendices A and B for details
on the calculation), and compared with the fit of the
ZEUS $F_2$
data in bins of $x$ to the form $a+blnQ^2$
(see reference \cite{caldw} and references therein for
more details on the data and the experimental fit).
The dotted line is the theoretical result obtained
with the pure CKMT model, and the bold (solid) line is the 
result obtained with
the NLO QCD-evoluted CKMT model when one takes $Q^2_0=2. GeV^2$
($Q^2_0=4. GeV^2$).

\vspace{0.5cm}

{\bf Figure 4.} 
$dlnF_2/dln(1/x)$ as a function of $Q^2$ calculated
by performing the NLO QCD evolution of the CKMT model, and 
compared to the fit
$F_2=Ax^{-\Delta_{eff}}$ of the ZEUS \cite{caldw}
and the E665 \cite{e665} data with $x<0.01$.
For details on the CKMT calculation, see Appendices A and B.
The dotted line is the theoretical result obtained
with the pure CKMT model, and the bold (solid) line is the result 
obtained with
the NLO QCD-evoluted CKMT model when one takes $Q^2_0=2. GeV^2$
($Q^2_0=4. GeV^2$).

\vspace{0.5cm}

{\bf Figure 5.} 
$dlnF_2/dln(1/x)$ as a function of $Q^2$ calculated
by performing the NLO QCD evolution of the CKMT model, and 
compared to the fit
$F_2=Ax^{-\Delta_{eff}}$ of the H1
data \cite{H1caldwell}.
For details on the CKMT calculation, see Appendices A and B.
The dotted line is the theoretical result obtained
with the pure CKMT model, and the bold (solid) line is the result 
obtained with
the NLO QCD-evoluted CKMT model when one takes $Q^2_0=2. GeV^2$
($Q^2_0=4. GeV^2$).

\vspace{0.5cm}

{\bf Figure 6.} 
Gluon density distribution as a function of $Q^2$ calculated 
by performing the NLO QCD evolution of
the CKMT model, and compared with the H1 Collaboration data
in reference \cite{adloff}. $g(x, Q^2)$ is plotted 
and not $xg(x, Q^2)$, in order
to show more clearly the evolution with the scale. In the theoretical
calculation, the bold (solid) line has been obtained by taking
a value of $Q^2_0$ at the starting point of the QCD evolution,
$Q^2_0=2. GeV^2$ ($Q^2_0=4. GeV^2$).

\vspace{0.5cm}
\newpage

{\bf Figure 7.} 
Gluon densities at $\mu^2=25. GeV^2$ as a function of $x$ calculated 
by performing the NLO QCD evolution of
the CKMT model, and compared to those determined from H1 DIS data (black dots) 
and from
H1 photoproduction data (stars). Experimental data on $D^*$ meson cross-section
measurements are from references \cite{zhokin,adloff}. In the theoretical calculation, the solid
(dotted) line corresponds to a value of $Q^2_0$ at the starting point of the 
QCD evolution,
$Q^2_0=2. GeV^2$ ($Q^2_0=4. GeV^2$).

\vspace{0.5cm}

{\bf Figure 8.} 
Gluon density at $\mu^2=200. GeV^2$ as a function of $x$ calculated 
by performing the NLO QCD evolution of
the CKMT model, to be compared with that obtained from the analysis 
of the H1 di-jet data \cite{zhokin,wobisch}. In the theoretical
calculation, the solid (dotted) line has been obtained by taking
a value of $Q^2_0$ at the starting point of the QCD evolution,
$Q^2_0=2. GeV^2$ ($Q^2_0=4. GeV^2$).

\vspace{0.5cm}

{\bf Figure 9.} 
Prediction of the behavior of $xg(x,Q^2)$ as a function of $x$ 
for several values of $Q^2$ measured both by H1 and ZEUS collaborations. 
The experimental points are not shown since the analysis of the more 
recent data is not completed. The solid (dotted) lines have been obtained by taking
a value of $Q^2_0$ at the starting point of the QCD evolution,
$Q^2_0=2. GeV^2$ ($Q^2_0=4. GeV^2$).
 
\end{document}